\documentclass[
    ,final            
  ]
  {aipproc}

\layoutstyle{6x9}

\newcommand{\be}{\begin{equation}}
\newcommand{\ee}{\end{equation}}
\newcommand{\ba}{\begin{eqnarray}}
\newcommand{\ea}{\end{eqnarray}}

\begin{document}

\title{Diffusion Processes in Turbulent Magnetic Fields}

\classification{95.30.Qd, 96.50.Tf, 98.38.-j, 98.38Gt }
\keywords      {Turbulence, MHD, Interstellar Medium, Plasma}

\author{A. Lazarian}{
  address={Department of Astronomy, University of Wisconsin-Madison, lazarian@astro.wisc.edu}
}

\begin{abstract}
 We study  of the effect of turbulence on diffusion processes within 
magnetized medium. While we exemplify our treatment with
 heat transfer processes, our
results are quite general and are applicable to different processes,
e.g. diffusion of heavy elements. Our treatment is also applicable 
to describing the diffusion of cosmic rays arising from magnetic field
wandering. 
In particular, we find that when the energy injection velocity is smaller
than the Alfven speed the heat transfer is partially suppressed,
while in the opposite regime the effects of turbulence depend on the 
intensity of driving.
In fact, the scale $l_A$ at which the turbulent velocity is equal the 
Alfven velocity is a new important parameter. When the electron mean
free path $\lambda$ is larger than $l_A$, the stronger the the turbulence, the
lower thermal conductivity by electrons is. The turbulent motions,
however, induces their own advective transport, that can provide
effective diffusivity. For clusters of galaxies, we find that the turbulence is
the most important agent for heat transfer. We also show that the domain of
applicability of the subdiffusion concept is rather limited.   
\end{abstract}

\maketitle


\section{Astrophysical Turbulence}

The magnetohydrodynamic (MHD) approximation is widely used to describe
the actual magnetized plasma turbulence over scales that are much
larger than both the mean free path of the particles and their Larmor
radius (see Kulsrud 2004). The theory of MHD turbulence has become
testable recently due to numerical simulations (see Biskamp 2003) 
which confirm (see Cho \& Lazarian 2005 and
ref. therein) the prediction of magnetized Alfv\'enic eddies
being elongated along magnetic field (see Shebalin, Matthaeus \&
Montgomery 1983, Higdon 1984) and provided results consistent with the
quantitative relations for the degree of eddy
elongation obtained  in Goldreich \& Sridhar (1995, henceforth GS95).

GS95 model assumes the isotropic injection of energy at scale
$L$ and the injection velocity equal to the Alfv\'en velocity in
the fluid $V_A$, i.e. the Alfv\'en Mach number
$M_A\equiv (\delta V/V_A)=1$.  This model can be easily generalized
for both $M_A>1$ and $M_A<1$ at the injection.  Indeed, if $M_A>1$,
instead of the driving scale $L$ for  one can use another scale,
namely $l_A$, which is the scale at
which the turbulent velocity gets equal to $V_A$.  For $M_A\gg 1$
magnetic fields are not dynamically important at large scales and the
turbulence follows the incompressible
 Kolmogorov cascade $V_l\sim l^{1/3}$ over the
range of scales $[L, l_A]$. This provides $l_A\sim L_D M_A^{-3}$.
If $M_A<1$, the turbulence obeys GS95 scaling (also called ``strong''
MHD turbulence) not from the scale $L$, but from a smaller scale $l_{trans}
\sim L M_A^{2}$ (Lazarian \& Vishniac 1999), while in the range
$[L, l_{trans}]$ the turbulence is ``weak''.  

\section{Basics of Diffusion in Magnetized Plasmas}

The issue of diffusion in magnetized plasma has been mostly dealt with in the 
context of heat transfer. Heat transfer in turbulent magnetized plasma is an important astrophysical 
problem which is relevant to the wide variety of circumstancies from mixing layers 
in the Local Bubble (see Smith \& Cox 2001) and Milky way 
(Begelman \& Fabian 1990) to cooling flows in intracluster medium (ICM) 
(Fabian 1994). The latter problem has been subjected to particular scrutiny
as  observations
do not support the evidence for the cool gas (see Fabian et al. 2001).
This is suggestive of the existence of heating that replenishes the 
energy lost via X-ray emission. Heat transfer from hot outer regions is
an important process to consider in this context.

It is well known that magnetic fields can 
suppress thermal conduction
perpendicular to their direction. The issue of heat transfer
 in realistic turbulent magnetic fields has been
long debated (see Bakunin 2005 and references therein). An influencial
paper by Narayan \& Medvedev (2001, henceforth NM01) obtained estimates of 
thermal 
conductivity by electrons using the GS95 model of MHD 
turbulence with 
the velocity $V_L$ at the energy 
injection scale $L$ that is equal to the Alfven velocity $V_A$, i.e.
the turbulence with the Alfven Mach number $M_A\equiv (V_L/V_A)=1$. 
This is rather
restrictive, as in the ICM $M_A>1$, while in other
astrophysical situations $M_A<1$.
Below we discuss turbulence for both $M_A>1$ and $M_A<1$ and compare
the particle diffusion to that by turbulent
fluid motions. For heat transfer, we refer to our results in Lazarian 
(2006, 2007).

Let us initially disregard the dynamics of fluid motions 
on diffusion, i.e. consider diffusion induced by particles
moving along static magnetic fields. 
Magnetized  turbulence in the GS95 model is anisotropic with
eddies elongated along (henceforth denoted by $\|$) 
the direction of local magnetic field. Consider
isotropic injection of energy at the outer scale $L$ and 
dissipation at the scale  $l_{\bot, min}$, where $\bot$ denotes the
direction of perpendicular to the local magnetic
field. NM01 observed that the separations of magnetic
field lines for $r_0<l_{\bot, min}$ are mostly influenced by
the motions at the scale $l_{\bot, min}$, which results in Lyapunov-type growth:
$\sim r_0 \exp(l/l_{\|, min})$. This growth is
similar to that obtained in earlier models with a single scale of turbulent
motions (Rechester \& Rosenbluth 1978, Chandran \& Cowley 1998). This is
not surprising as the largest shear that causes field line divergence
is provided by the marginally damped motions at the scale around $l_{\bot, min}$.
In NM01 $r_0$ is associated with the size of the
cloud of electrons of the electron
Larmor radius $r_{Lar, particle}$. They find that the electrons
should travel over the distance
\begin{equation} 
L_{RR}\sim l_{\|, min} \ln (l_{\bot, min}/r_{Lar, particle})
\label{RR}
\end{equation}
 to get separated by $l_{\bot, min}$.

 Within the single-scale model which formally corresponds to
$L=l_{\|, min}=l_{\bot, min}$ the scale $L_{RR}$
is called Rechester-Rosenbluth distance. For the ICM parameters the 
logarithmic factor in Eq. (\ref{RR}) is
of the order of $30$, and this causes $30$ times decrease of thermal
conductivity for the single-scale models\footnote{For the single-scale model
$L_{RR}\sim 30L$ and the diffusion over distance $\Delta$ takes 
$L_{RR}/L$ steps, i.e. $\Delta^2\sim L_{RR} L$, which decreases the
corresponding diffusion coefficient $\kappa_{particle, single}\sim \Delta^2/\delta t$ by the factor of 30.}.  In the multi-scale models
with a limited (e.g. a few decades) inertial range
the logarithmic factor stays of the
same order but it does not affect the thermal conductivity,
provided that $L\gg l_{\|, min}$. 
Indeed, for the electrons to diffuse isotropically they
should spread from $r_{Lar,particle}$ to $L$. The GS95
model of turbulence operates with field lines that are sufficiently stiff,
i.e. the deviation of the field lines from their original direction is
of the order unity at scale $L$ and less for smaller scales.
Therefore to get separated
from the initial 
distance of $l_{\bot, min}$ to a distance $L$ (see Eq. (\ref{subA}) with $M_A=1$), at which the motions get 
uncorrelated the electron should diffuse the distance slightly larger (as 
field lines are not straight) than 
$\sqrt{2}L$, which is much larger than the extra 
travel distance 
$\sim 30 l_{\|, min}$. Explicit calculations in NM01 support this intuitive
picture.

\section{Diffusion for $M_A>1$}

Turbulence with $M_A>1$ evolves along hydrodynamic
isotropic Kolmogorov cascade, i.e. $V_l\sim V_L (l/L)^{1/3}$ over the
range of scales $[L, l_A]$, where 
\begin{equation}
l_A\approx  L (V_A/V_L)^{3}\equiv L M_A^{-3},
\label{lA}
\end{equation}
is the scale at which the magnetic field gets dynamically important, i.e.
 $V_l=V_A$. This scale plays the role
of the injection scale for the GS95 turbulence, i.e. $V_l\sim V_A (l_{\bot}/l_A)^{1/3}$,
with eddies at scales less than $l_A$ 
geting elongated in the direction of the local magnetic field. 
The corresponding anisotropy can be characterized by the relation between
the semi-major axes of the eddies 
\begin{equation}
l_{\|}\sim L (l_{\bot}/L)^{2/3} M_A^{-1},~~~ M_A>1,
\label{supA}
\end{equation}
 where
$\|$ and $\bot$ are related to the direction of the local magnetic field. 
In other words, for $M_A>1$, the turbulence is still isotropic at the 
scales larger to $l_A$, but 
develops $(l_{\bot}/l_A)^{1/3}$ anisotropy for $l<l_A$.

If particles (e.g. electrons) mean free path 
$\lambda\gg l_A$, they stream freely over the distance of $l_A$. 
For particles initially at distance $l_{\bot, min}$ to get separated by $L$ the
required travel is the random walk with the step $l_A$, i.e. the mean-squared
displacement  of a particle till it enters an independent large-scale 
eddy  
$\Delta^2\sim l_A^2 (L/l_A)$, where $L/l_A$ is the number of steps.
These steps require time $\delta t\sim (L/l_A) l_A/C_1v_{particle}$,
where $v_{particle}$ is electron thermal velocity and the coefficient $C_1=1/3$ 
accounts for 1D character of motion along 
 magnetic field lines. Thus
the electron diffusion coefficient is
\begin{equation}
\kappa_{particle}\equiv \Delta^2/\delta t\approx (1/3) l_A v_{particle},~~~ l_A<\lambda,
\label{el}
\end{equation}
which for $l_A\ll \lambda$ constitutes a substantial reduction of diffusivity
compared to its unmagnetized value $\kappa_{unmagn}=\lambda v_{particle}$.
We assumed in Eq. (\ref{el}) that $L\gg 30 l_{\|, min}$ (see \S 2.1).

 For $\lambda\ll l_A\ll L$,  $\kappa_{particle}\approx 1/3 \kappa_{unmagn}$ as both the $L_{RR}$
and the additional distance for electron to diffuse because of magnetic
field being stiff at scales less than $l_A$ are negligible compared
to $L$. For $l_A\rightarrow L$, when magnetic field has rigidity
up to the scale $L$, it gets around $1/5$ of
the value in unmagnetized medium, according to NM01.

Note, that even dynamically unimportant magnetic
fields do influence heat conductivity over short time intervals.
For instance, over time interval less than
$l_A^2/C_1\kappa_{unmagn}$ 
 the diffusion happens along stiff magnetic field lines 
and the difference between parallel and perpendicular 
diffusivities is large\footnote{The relation between the
mean squared displacements perpendicular to magnetic field 
$\langle  y^2\rangle$ and the displacements $x$ along magnetic field
for $x<l_A$ can be obtained through the diffusion equation
approach in \S 2.3 and
Eq.~(\ref{supA}). This gives $\langle  y^2\rangle^{1/2} \sim \frac{x^{3/2}}{3^{3/2}L^{1/2}} M_A^{3}$.}.
This allows the transient existence of sharp small-scale temperature gradients.

\section{Diffusion for $M_A<1$}

 It is intuitively clear that
for $M_A<1$ turbulence should be anisotropic from the injection scale $L$.
In fact, at large scales the turbulence is expected to be  {\it weak}\footnote{The terms ``weak'' and ``strong'' turbulence are accepted in the literature, but
can be confusing. As we discuss later at smaller scales at which the turbulent
velocities decrease the turbulence becomes {\it strong}. The formal theory of
weak turbulence is given in Galtier et al. (2000).} 
(see Lazarian \& Vishniac 1999, henceforth LV99). 
Weak turbulence is characterized
by wavepackets that do not change their $l_{\|}$, but develop structures 
perpendicular to magnetic field, i.e. decrease $l_{\bot}$ . This cannot
proceed indefinitely, however. At some small scale 
 the GS95 condition of {\it critical balance}, i.e. $l_{\|}/V_A\approx l_{\bot}/V_l$, becomes satisfied. This perpendicular scale $l_{trans}$ 
can be obtained substituting the scaling of
weak turbulence (see LV99) $V_l\sim V_L(l_{\bot}/L)^{1/2}$  into
the critical balance condition.
This provides $l_{trans}\sim L M_A^2$ and the corresponding
velocity $V_{trans}\sim V_L M_A$. For scales less than $l_{trans}$
the turbulence is {\it strong} and it follows the scalings of the
GS95-type, i.e. $V_l\sim V_L(L/l_{\bot})^{-1/3} M_A^{1/3}$ and
\begin{equation}
 l_{\|}\sim L (l_{\bot}/L)^{2/3} M_A^{-4/3},~~~ M_A<1.
\label{subA}
\end{equation}

For $M_A<1$, magnetic field wandering in the direction
perpendicular
to the mean magnetic field (along y-axis) can be described
by $d\langle  y^2\rangle/dx \sim \langle  y^2\rangle/l_\|$ (LV99),
where\footnote{The fact that
one gets $l_{\|, min}$ in Eq. (\ref{RR}) is related to the presence of this
scale in this diffusion equation.} $l_\|$ is expressed by Eq. (\ref{subA})
and one can associate $l_{\bot}$ with
$2\langle  y^2\rangle$
\begin{equation}
 \langle  y^2\rangle^{1/2} \sim \frac{x^{3/2}}{3^{3/2}L^{1/2}}
M_A^{2},~~~ l_{\bot}<l_{trans}
\label{3}
\end{equation}
For 
weak turbulence $d\langle  y^2\rangle/dx \sim LM_A^4$ (LV99) and thus 
\begin{equation}
\langle  y^2\rangle^{1/2} \sim L^{1/2} x^{1/2} M_A^2,~~~ l_{\bot}>l_{trans}.
\label{weak}
\end{equation}  

Fig.~1 confirms the correctness of the above scaling numerically.

\begin{figure*}
  \includegraphics[width=\textwidth]{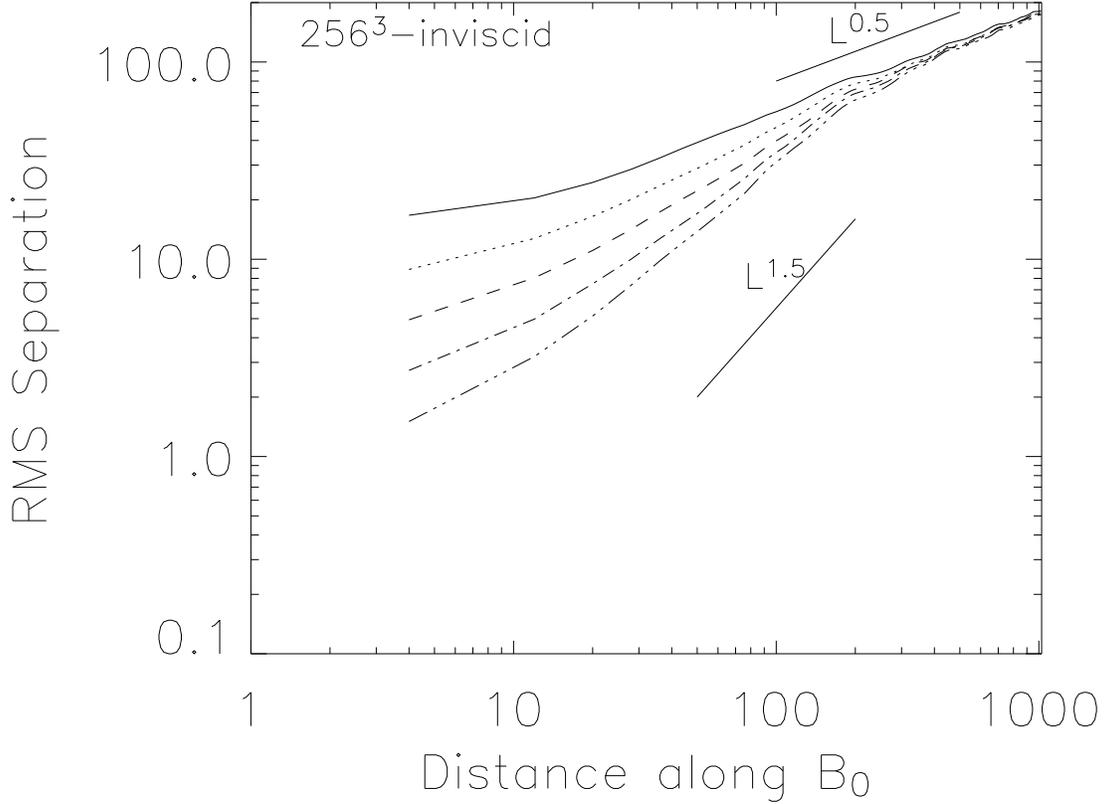}
\caption{Root mean square separation of field lines in a simulation of
inviscid MHD turbulence, as a function of
distance parallel to the mean magnetic field, for a range of initial
separations.  Each curve represents 1600 line pairs.
The simulation has been filtered to remove pseudo-Alfv\'en
modes, which introduce noise into the diffusion calculation. From Lazarian,
Vishniac \& Cho 2004.}
\end{figure*}

Eq.~(\ref{3}) differs by the factor $M_A^{2}$ from that
in NM01, which reflects the gradual suppression of 
thermal conductivity perpendicular to the mean magnetic field
as the magnetic field gets stronger. Physically this means that for
$M_A<1$ the magnetic field fluctuates around the well-defined
mean direction. Therefore the diffusivity gets anisotropic
with the diffusion coefficient parallel to the mean field 
$\kappa_{\|, particle}\approx 1/3 \kappa_{unmagn}$ being larger than coefficient
for diffusion perpendicular to magnetic field $\kappa_{\bot, particle}$.

Consider the coefficient $\kappa_{\bot, particle}$ for $M_A\ll 1$. As  
NM01 showed, particles become uncorrelated if they are displaced over the
 distance $L$ in the direction perpendicular to
 magnetic field.  To do this, a particle has first
to travel 
$L_{RR}$ (see Eq.~(\ref{RR})),
 where Eq.~(\ref{subA}) relates $l_{\|, min}$ and $l_{\bot, min}$.
 Similar to the case in \S 2.1, for
$L\gg 30 l_{\|, min}$, the additional travel arising from the
 logarithmic factor is 
negligible compared to the overall diffusion distance $L$.
 At larger scales electron has to diffuse 
$\sim L$ in the direction parallel to magnetic field to cover the distance of
$L M_A^2$ in the direction
 perpendicular to magnetic field direction. To diffuse over a 
distance R with random walk of
$LM_A^2$ one requires $R^2/L^2M_A^4$ steps. The time of the individual step
is $L^2/\kappa_{\|, particle}$. Therefore the perpendicular diffusion 
coefficient is 
\begin{equation}
\kappa_{\bot, particle}=R^2/(R^2/[\kappa_{\|, particle}M_A^4])=\kappa_{\|, particle}M_A^4,~~~ M_A<1,
\label{9}
\end{equation} 
An essential assumption there is
that the particles do not trace their way back over the individual steps along 
magnetic field lines, i.e. $L_{RR}<<L$. Note, that
for $M_A$ of the order of unity this is not accurate and
 one should account for the actual 3D displacement. This introduces the change
by a factor of order unity (see above).

\section{Turbulent Diffusivity}

Turbulent motions themselves can induce advective
 transport. In Cho et al. (2003) we dealt with the turbulence
with $M_A\sim 1$ and estimated
\begin{equation}
\kappa_{dynamic}\approx C_{dyn} L V_L,~~~ M_A>1,
\label{dyn}
\end{equation}
where $C_{dyn}\sim 0(1)$ is a constant, which for hydro turbulence
is around $1/3$ (Lesieur 1990). 
If we deal with heat transport, for fully ionized non-degenerate plasma we assume $C_{dyn}\approx 2/3$ to account for the advective heat transport by 
both protons and electrons\footnote{This gets clear if one uses the 
heat flux equation
$q=-\kappa_c\bigtriangledown T$, where $\kappa_c=nk_B\kappa_{dynamic/electr}$,
$n$ is {\it electron} number density, and $k_B$ is the Boltzmann constant,
for both electron and advective heat transport.}.
Thus eq.~(\ref{dyn}) covers the cases of
both $M_A>1$ up to $M_A\sim 1$. For $M_A<1$
one can estimate $\kappa_{dynamic}\sim d^2\omega$, where $d$ is
the random walk of the field line over the wave period $\sim \omega^{-1}$. 
As the weak turbulence
at scale $L$ evolves over time $\tau\sim 
M_A^{-2}\omega^{-1}$, $\langle y^2 \rangle$  is the result of
the random walk with a step $d$, i.e.
$\langle y^2 \rangle\sim (\tau\omega)d^2$. According to eq.(\ref{3}) and
(\ref{weak}), the  field line is displaced over  time $\tau$ by
$\langle y^2\rangle\sim L M_A^4 V_A \tau$. Combining the two one gets
 $d^2 \sim L M_A^3 V_L \omega^{-1}$, which provides 
$\kappa_{dynamic}^{weak}\approx C_{dyn} LV_L M_A^3$, 
which is similar to the diffusivity arising from strong turbulence at
scales less than $l_{trans}$, i.e. $\kappa_{dynamic}^{strong}\approx C_{dyn} l_{trans}
V_{trans}$. The total diffusivity is the sum of the two, i.e. for
plasma
\begin{equation}
\kappa_{dynamic}\approx (\beta/3) LV_L M_A^3, ~~~ M_A<1,
\label{dyn_weak}
\end{equation}
where $\beta\approx 4$.

\section{Example: Thermal Conductivity}

In thermal plasma, electrons are mostly responsible for thermal conductivity.
The schematic of the parameter space for  $\kappa_{particle}<\kappa_{dynamic}$ is shown in Fig~2, where the 
the Mach number $M_s$ 
 and the Alfven Mach number $M_A$ are 
the variables. For $M_A<1$, the ratio of diffusivities arising from
fluid and particle motions is
 $\kappa_{dynamic}/\kappa_{particle}\sim \beta\alpha M_S M_A(L/\lambda)$ (see 
Eqs. (\ref{9}) and (\ref{dyn_weak})), the square root of the
ratio of the electron to proton mass $\alpha=(m_e/m_p)^{1/2}$, which
provides the separation line between the two regions in Fig. 2, $\beta\alpha M_s\sim
(\lambda/L) M_A$. For $1<M_A<(L/\lambda)^{1/3}$ the mean free path is less
than $l_A$ which results in $\kappa_{particle}$ being some fraction of
$\kappa_{unmagn}$, while $\kappa_{dynamic}$ is  given by Eq.~(\ref{dyn}).
Thus $\kappa_{dynamic}/\kappa_{particle}\sim \beta \alpha M_s (L/\lambda)$, i.e. the ratio
 does not depend on $M_A$ (horisontal line in Fig.~2). When $M_A>(L/\lambda)^{1/3}$ the
mean free path of electrons is constrained by $l_A$. In
this case $\kappa_{dynamic}/\kappa_{particle}\sim \beta \alpha M_s M_A^3$ (see 
Eqs. (\ref{dyn})
and (\ref{el})) . This results in the separation line 
$\beta\alpha M_s \sim M_A^{-3}$
in Fig.~2.

\begin{figure}
\hbox{
  \includegraphics[height=.45\textheight]{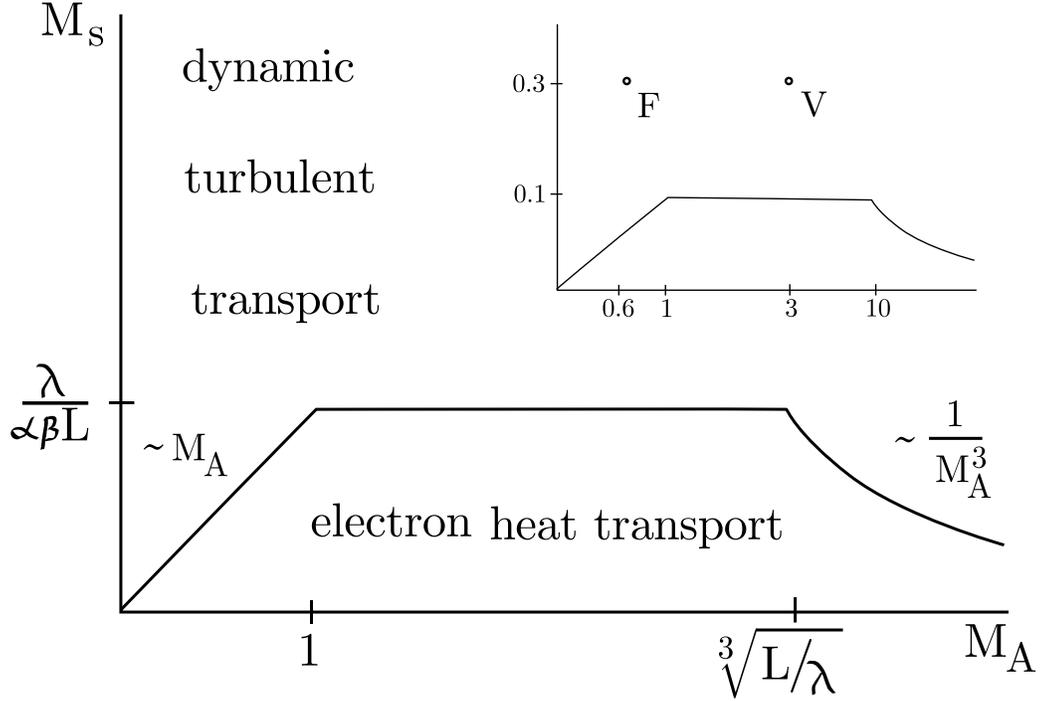}
}
  \caption{Parameter space for particle diffusion or turbulent diffusion to
dominate: application to heat transfer. Sonic Mach number $M_s$ is ploted
against the Alfven Mach number $M_A$. The heat transport is dominated by the 
dynamics of turbulent eddies is above the curve and by thermal conductivity of
 electrons is below the curve. Here $\lambda$ is the mean free path of the
electron, $L$ is the driving scale, and $\alpha=(m_e/m_p)^{1/2}$, $\beta\approx 4$. The panel in the right upper coner of the figure illustrates 
heat transport for the parameters for a cool
core Hydra cluster (point ``F''), ``V'' corresponds
to the illustrative model of a cluster core in Ensslin et al. (2005). From
Lazarian (2007).}
\end{figure}

\section{Heat Transfer in Intracluster Medium}

It is generally believed that Intracluster Medium (ICM) is turbulent.
The considerations below can be used as guidance.
  In unmagnatized
plasma with the ICM temperatures $T\sim 10^8$ K and and density $10^{-3}$
cm$^{-3}$ 
the kinematic viscosity $\eta_{unmagn} \sim v_{ion} \lambda_{ion}$, 
where $v_{ion}$ and $\lambda_{ion}$ are the velocity of an ion and
its  mean free path, respectively, would make the Reynolds number 
$Re\equiv LV_L/\eta_{unmagn}$ of the order of 30. This is barely enough for the onset
of turbulence. For the sake of simplicity we assume that ion mean free path coinsides with the proton mean free path and both scale as $\lambda\approx
3 T_3^2 n_{-3}^{-1}$~kpc, where the temperature $T_3\equiv kT/3~{\rm keV}$ and
$n_{-3}\equiv n/10^{-3}~{\rm cm^{-3}}$. This provides
$\lambda$  of the order of 0.8--1 kpc for
the ICM (see NM01).   

It is accepted, however, that magnetic fields decrease the diffusivity. 
Somewhat naively assuming the maximal
scattering rate of an ion, i.e. scattering every orbit (the
so-called Bohm diffusion limit) one gets the viscosity perpendicular
to magnetic field $\eta_{\bot}\sim v_{ion} r_{Lar, ion}$, which is much smaller
than $\eta_{unmagn}$, provided that the ion Larmor radius $r_{Lar, ion}\ll \lambda_{ion}$.
For the parameters 
of the ICM this allows essentially invicid motions\footnote{A {\it regular} magnetic field
 $B_\lambda\approx (2 m k T)^{1/2} c/(e\lambda)$ that makes
$r_{Lar, ion}$ less than $\lambda$ and therefore $\eta_{\bot}<\nu_{unmagn}$
is just $10^{-20}$~G. {\it Turbulent} magnetic field with many reversals 
over $r_{Lar, ion}$ does not interact efficiently with a proton, however. As the
result, the protons are not constrained 
until $l_A$ gets of the order of $r_{Lar, ion}$. This happens when 
the turbulent magnetic field is of the
 order of $2\times 10^{-9}(V_L/10^3{\rm km/s})$~G. At this point, the step for
the random walk is $\sim 2\times 10^{-6}$~pc and the Reynolds
number is $5\times 10^{10}$. }
 of magnetic lines 
parallel to each other, e.g. Alfven motions.

In spite of the substantial progress in understading  of the ICM
 (see En{\ss}lin, Vogt \& Pfrommer 2005, henceforth EVP05, En{\ss}lin \&
Vogt 2006, henceforth EV06 
and references therein), the basic parameters of ICM
turbulence are known within the factor of 3 at best. For instance, the
estimates of  injection velocity $V_L$ varies in the literature
from 300 km/s to $10^3$ km/s,
while the injection scale $L$ varies from 20 kpc to 200 kpc, 
depending whether
the injection of energy by galaxy mergers or galaxy wakes is considered. 
EVP05 considers an {\it illustrative} model in which the magnetic field
 with the 
10 $\mu$G fills 10\% of
the volume, while 90\% of the volume is filled with the field of $B\sim 1$
 $\mu$G. Using the latter number and  assuming
 $V_L=10^3$ km/s, $L=100$ kpc, and the density of the hot ICM
is $10^{-3}$ cm$^{-3}$, one gets $V_A\approx 70$ km/s, i.e. $M_A>1$.
 Using the numbers above, one
 gets $l_A\approx 30$ pc for the 90\% of the volume of the hot ICM,
which is much less than $\lambda_{ion}$. The diffusivity of ICM plasma
gets $\eta=v_{ion} l_A$ which for the parameters above provides
 $Re\sim 2\times 10^3$, which is enough for driving superAlfvenic turbulence
at the outer scale $L$. However, as $l_A$ increases as $\propto B^3$, $Re$ gets around $50$ for
the  field of 4 $\mu$G, which
is at the border line of exciting turbulence\footnote{One can imagine
dynamo action in which superAlfvenic turbulence generates magnetic
field till $l_A$ gets large enough to shut down the turbulence.
}. However, the regions with higher
magnetic fields (e.g. 10 $\mu$G)
can support Alfvenic-type turbulence with the injection scale $l_A$ and
the injection velocities resulting from large-scale shear
$V_L (l_A/L)\sim V_L M_A^{-3}$.

For the regions of $B\sim 1$ $\mu$G the value of $l_A$ is
smaller than the mean free path of electrons $\lambda$.
 According to Eq.~(\ref{el}) the value of
$\kappa_{electr}$ is 100 times smaller than $\kappa_{Spitzer}$. 
On the contrary, $\kappa_{dynamic}$ for the ICM parameters adopted will be
$\sim 30 \kappa_{Spitzer}$, which makes 
 the dynamic diffusivity the dominant process. This agrees well 
with the observations in Voigt \& Fabian (2004). Fig.~2 shows the dominance
of advective heat transfer for the parameters 
of the cool core of Hydra A ( $B=6$~$\mu$G,
$n=0.056$ cm$^{-3}$, $L=40$~kpc, $T=2.7$~keV according to 
EV06), point ``F'', and for the illustrative model in EVP05, point ``V'',
for which $B=1$~$\mu$G.

Note that our stationary model
of MHD turbulence is not directly applicable to transient
wakes behind galaxies.
 The ratio of the
damping times of the hydro turbulence and the time of
 straightening of the magnetic field
lines is $\sim M_A^{-1}$. Thus, for $M_A>1$, the magnetic field at scales
larger than $l_A$ will be straightening 
gradually after the hydro turbulence has faded away over time $L/V_L$. 
The process can be characterized as injection of turbulence at velocity
$V_A$ but at scales that increase linearly with time, i.e. as $l_A+V_At$.
 The
study of heat transfer in transient turbulence and
 magnetic field ``regularly'' stretched by passing galaxies
 will be provided elsewhere.

\section{Aspects of Cosmic Ray Diffusion}

Diffusion of cosmic rays (CR) includes diffusion arising from CR scattering and
that of wandering magnetic field lines. For our purposes, we disregard the
diffusion perpendicular to magnetic field lines that arises from gyroresonance
interactions. 

Naturally, the diffusion of magnetic field lines depends on the adopted model of turbulence. For instance, Lazarian \& Beresnyak (2006) discuss a mechanism by which slab-type perturbations can be generated on the scale of CR gyroradius (seeFig. 3). These slab-type perturbations are likely to contribute subdominantly
to field line wandering. Therefore below we concentrate, as in the rest of
the paper, on the field wandering induced by Alfvenic modes of GS95 turbulence.

\begin{figure}
\hbox{
  \includegraphics[height=.45\textheight]{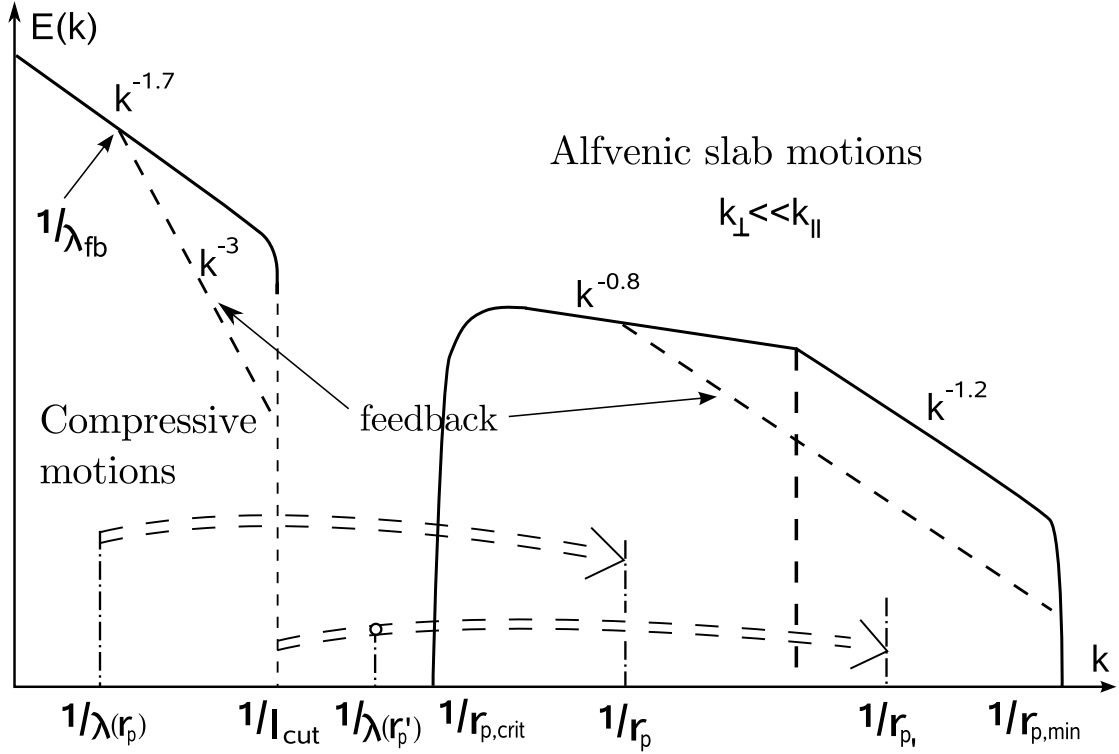}
}
  \caption{Energy density of compressive modes and Alfv\'enic slab-type
waves, induced by CRs. The energy is transferred from
the mean free path scale to the CR Larmor radius scale.
If the mean free path falls below compressive motions cutoff or feedback
suppression scale, the spectrum of slab waves becomes steeper. From
Lazarian \& Beresnyak (2006). }
\label{fig1}
\end{figure}

Generalizing our arguments above in terms of the diffusion parallel and
perpendicular to mean magnetic field, one can write that 
$\kappa_{\bot, cr}=\kappa_{\|, cr}(\delta R)^2/(\delta z)^2$, where $\delta R$
is a random walk step in the direction perpendicular to magnetic field. 
If the corresponding scale $\delta z$ is less than $L_{RR}$, the motion
along the field line is one dimentional diffusion that retraces its steps.
In this case, $\delta z=(\kappa_{\|, cr} \delta t)^{1/2}$. If, following 
earlier authors (see Kota \& Jokipii 2000, Web et al. 2006) we introduce a spatial field diffusion coefficient
$\kappa_{spat}=(\delta R)^2/(\delta z)$, we easily get
$\kappa_{\bot, cr}\approx \kappa_{spat}\kappa_{\|, cr}^{1/2}(\delta t)^{1/2}$,
which results in the {\it subdiffusion} in perpendicular direction, i.e.
in the distance perpendicular to magnetic field growing as 
$\kappa_{\bot, cr}\delta t=\kappa_{spat} \kappa_{\|, cr}^{1/2} (\delta t)^{1/2}$.
However, as soon as the distance to diffuse is much larger than $L_{RR}$ (see
Eq.(\ref{RR}), the
subdiffusive effects are negligible.

Although CR velocities are of the order of light speed, for sufficiently small
mean free paths the advection of CR by turbulent motions may become important. 

 \section{Concluding Remarks}

In the paper above we attempted to describe the diffusion by 
particle and turbulent motions for $M_A<1$ and $M_A>1$. 
Unlike earlier papers, we find that turbulence may both enhance diffusion
 and suppress it. For instance, when
$\lambda$ gets larger than $l_A$ the 
conductivity of the medium $\sim M_A^{-3}$ and therefore the turbulence 
{\it inhibits} heat transfer, provided that
$\kappa_{particle}>\kappa_{dynamic}$. Along with the plasma effects that we mention
below, this effect can, indeed, support
sharp temperature gradients in  hot plasmas
with weak magnetic  field.

As discussed above, rarefied plasma, e.g. ICM plasma,
 has large viscosity for motions parallel
to magnetic field and marginal viscosity for motions that induce
perpendicular mixing. Thus fast dissipation of sound waves in the ICM 
does not contradict the medium being turbulent. The later may be 
important for the heating of central regions of clusters caused by
 the AGN feedback
(see Churasov et al. 2001, Nusser, Silk \& Babul 2006 and more 
references in EV06). Note, that 
 models that include both heat transfer from the outer hot regions and an additional heating from the AGN feedback
look rather promissing (see Ruszkowkski \& Begelman 2002, 
Piffaretti \& Kaastra 2006).
We predict that the viscosity for
1 $\mu$G regions is
less than for 10 $\mu$G regions and therefore heating by sound waves (see Fabian
et al. 2005) could be more efficient for the latter. 
Note, that the plasma instabilities in collisionless magnetized ICM 
arising from compressive motions (see Schekochihin \& Cowley 2006, 
Lazarian \& Beresnyak 2006) can resonantly scatter particles
and decrease $\lambda$. 
This decreases further $\kappa_{particle}$ compared to $\kappa_{unmagn}$ but
increases $Re$.   
 In addition, we disregarded mirror effects that can reflect electrons
back (see Malyshkin \& Kulsrud 2001 and references therein),
which can further decrease $\kappa_{particle}$.

All in all, we have shown that it is impossible to characterize the
diffusion in magnetized plasma by a single fraction of the diffusion
coefficient of unmagnetized
value. The diffusion depends on sonic and
Alfven Mach numbers
of turbulence and the corresponding diffusion coefficient 
may be much higher and much lower than the unmagnetized one.
As the result, turbulence can inhibit or enhance diffusivity depending
on the plasma magnetization and turbulence driving.

 Our study indicates that
in many cases related to ICM the advective heat transport by dynamic turbulent eddies 
dominates thermal conductivity. In addition, ``subdiffusivity''
is probably subdominant for many astrophysical problems.


{\bf Acknowledgments}  The research is supported by
 the NSF Center for Magnetic Self Organization in
Laboratory and Astrophysical Plasmas. Help by A. Beresnyak is acknowledged.


\end{document}